\documentclass[aps,prl,twocolumn,groupedaddress,showpacs]{revtex4}
\usepackage{graphicx}

\begin{document}
\title{Tailoring Single and Multiphoton Probabilities of a Single 
Photon On-Demand Source}
\author{A. L. Migdall}
\author{D. Branning}
\altaffiliation{now at University of Illinois at Urbana-Champaign}
\author{S. Castelletto}
\altaffiliation{also at Istituto Elettrotecnico Nazionale G. Ferraris}
\author{M. Ware}
\affiliation{Optical Technology Division, National Institute of Standards and Technology, Gaithersburg, 
Maryland 20899}

\date{\today}

\begin{abstract}
As typically implemented, single photon sources cannot be made to produce 
single photons with high probability, while simultaneously suppressing 
the probability of yielding two or more photons. Because of this, single photon
 sources cannot really produce single photons on demand. We describe a 
multiplexed system that allows the probabilities of producing one and more 
photons to be adjusted independently, enabling a much better approximation 
of a source of single photons on demand.
\end{abstract}


\maketitle
\parskip=0 pt

The advent of photon-based quantum cryptography, communication and computation 
schemes  \cite{BEB89, ERT92}
 is increasing the need for light sources that produce individual photons. It is of particular
 importance that these single photons be produced 
in as controlled a manner as possible, as unwanted additional photons can render quantum 
cryptographic links insecure and degrade quantum computation efficiencies 
\cite{BLM00}. 

Single photons (or more precisely, approximations thereof) are now commonly created 
via the process of parametric downconversion (PDC) \cite{BUW70, ERT92,   GRT01},
 although attenuated lasers and quantum-dots and other single quantum site 
 sources are also used \cite{JAF96,  KBK99, BHL00,
  YKS01}.
  Because PDC creates photons in pairs,
 the detection of one photon indicates, or ``heralds'', the existence of its twin, a significant
 advantage over other methods (even aside from the potential to directly produce entangled states). 
In addition, because the PDC process is governed by the constraints of 
phase matching, it is possible to know the output trajectory, polarization, and 
wavelength of that heralded photon.  While PDC has a long history as a single
 photon source and there have been many recent improvements \cite{KWW99}, the scheme
 has a couple of problems. The conversion process is random, so while an output
 photon is heralded by its twin, there is no control or prior knowledge of when the heralding event will
 occur. In addition, there is a possibility of producing more than one pair 
 at a time and because that 
probability increases nonlinearly with the one photon probability, one must operate at low one
 photon probabilities. So to be assured that more than one photon is not produced, one must 
operate where it is most likely that no photon is produced at all \cite{NIC00}.

The faint laser scheme suffers the same difficulty as the PDC method, in 
possibly producing more than one photon at a time, with the added difficulty of not having any herald 
at all \cite{BHK98}. Quantum-dot sources offer promise as a new way of definitively 
producing single photons, although it remains to be seen whether their output/collection 
efficiencies can be made 
to approach unity in practice, a requirement for a true on-demand source.

To surmount the problem of random production in PDC, one uses a pulsed laser to 
pump the nonlinear crystal (see for instance \cite{BHL00}). With a pulsed source,
 photon pairs can only be produced at certain times. Unfortunately the 
multiple photon emission problem remains; a high probability ($P_{1}$) of producing a single photon pair during
 each pulse leads to an increased likelihood ($P_{>1}$) of producing more than one photon pair during each pulse,
 defeating the original goal of having a source of single photons. This problem occurs regardless of whether
 sources with Poisson or Bose statistics are used \cite{GRT01,WAM95}.  Because of this,
 pulsed systems are usually operated with low probability of producing an output photon pair during a 
laser pulse ($P_1\sim $0.1 to 0.3) \cite{JAF96, BHK98, BHL00}. Thus, while photons can only come during
 specific time
 windows, most of these time windows will contain no photons at all, illustrating the trade off between 
the two requirements of producing a photon on demand and being assured that there is, in fact, just one photon. 

The improvement presented here allows these two competing requirements to be adjusted independently
 by decoupling $P_{1}$ and $P_{>1}$. We can then select both the desired likelihood of production 
of a photon pair and the desired suppression of multiple pair events. This is accomplished 
using an array of downconverters and detectors (Fig. 1a) pumped simultaneously
 by the same laser pulse. The laser power is chosen so each downconverter has some small
 probability of producing a photon pair, while the number of downconverters is chosen so there is a high
 likelihood of at least one pair being created somewhere in the array. The detector associated with each 
downconverter allows us to determine which of the downconverters has fired. This information is used to 
control an optical switching circuit directing the other photon of the pair onto the single 
output channel. This arrangement allows a much truer approximation of a single 
photon on-demand source than
 is possible with other methods.

A simple extension of this arrangement can produce a regularly spaced series of single
 photons. By ganging up a series of these ``single photon on demand 
 setups'' with additional optical switches
 and a series of optical delays, it is possible to produce an arbitrary length train of single photons. By producing
 a pulse train long enough to last until the next pump laser pulse arrives, 
 one can create a continuous
 train of single photon pulses synchronized with an external signal.

These basic concepts could be used to produce higher order photon number states as well. By using
 detectors with the capability of sensing the number of photons in a single pulse, the 
switching circuit could just as well direct the outputs of those converters that produced multiple photon
 pairs to the output channel. That would result in an output pulse train with each pulse containing the 
desired number of photons.
 \begin{figure}
 \includegraphics{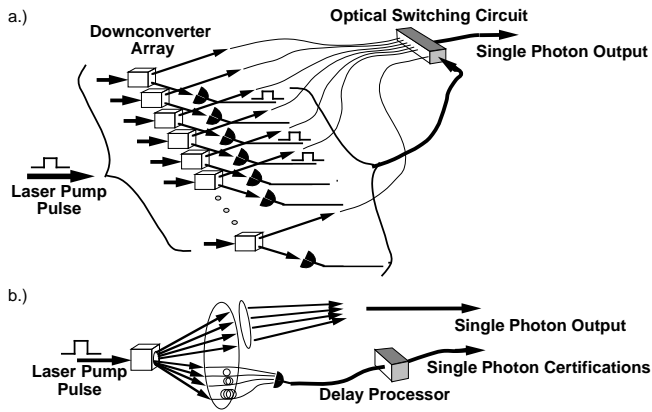}%
 \caption{\label{fig1} a.) Single photon source using an array of crystals 
 and trigger 
detectors. Which-trigger information is sent to the optical switching 
circuit. Input line delays to allow the trigger information to arrive before the incoming photons are
 not shown. b.) Simplified implementation using only a single downconverter crystal and detector and 
no optical switching circuit. Multiple staggered length delays are shown leading to the trigger 
detector. A lens collects all the modes correlated to the trigger.}
 \end{figure}

While the scheme just described is conceptually the 
simplest way of presenting the method, there are a number of modifications that can improve
 the efficiency, construction, and convenience of the system. The first of these is that the array of 
downconverters can be implemented with a single PDC crystal. 
This is possible because 
while  phasematching requires a PDC photon pair to be constrained to a plane
 containing the pump beam, the azimuthal angle is not constrained. Thus, 
 the PDC process 
produces light distributed azimuthally around the pump laser direction (for type I phasematching). So,
 each azimuthal plane can be thought of as a separate downconverter. Thus the multiple 
PDC setup is achieved by placing many detectors azimuthally in an annular pattern around
 the pump direction of a single downconverter and collecting at the correlated positions. 

A second modification allows the array of detectors to be replaced by a single detector. By sending
 each of the potential herald photons to an increasing sequence of delays and then directing the delay 
outputs to a single detector, the timing of the detector pulse indicates which of the input downconverters
 has created a pair. Of course, several of the downconverters may produce a photon pair, but only the first
 photon herald received by the detector causes it to fire. Detector 
 deadtime causes subsequent 
photons to be ignored. The timing of the detector pulse is used to select which correlated photon channel 
to direct to the output of system.  

A third implementation (Fig. 1b) even eliminates the output switching 
network circuit, while still maintaining a significant advantage over 
both the conventional PDC and faint laser
  photon sources.
 The output photons are simply collected with a single lens output port. 
This most basic implementation allows
 production of pulses with individual certainties of that pulse containing 
 exactly one photon. In
 other words, this source provides single photons and a ``certificate" tied to each output photon quantifying
 the likelihood that just one photon was emitted. Some of these pulses can achieve a significantly higher 
single photon certainty than is possible with the conventional single photon source setups. This more 
complete characterization of the emitted pulse and its tighter constraints on 
$P_{>1}$, will allow more efficient use of the 
emitted light. This reduces the need for overhead tasks such as privacy 
amplification \cite{ILM01}.

The basic reason that this arrangement can produce single photons with lower probability of multiple photons is that the delay system provides extra information about the photons produced. 
For instance,
 in the cases where one of the longer delays happened to cause the detector to fire, we know that all the 
prior delays did not cause the detector to fire. If the detection quantum efficiency is high, 
 it is very likely that there were no photons in those modes coupled to those 
 shorter paths. Thus $P_{>1}$ is greatly reduced because it is just the 
 multiphoton probability for only the last delay, rather than for all $N_D$ of 
 them,  which has a mean photon number of $\overline{n}$ vs. $\overline{n}/N_D$.
 
We now quantify the advantage of this last scheme in its simplest and most 
straightforward implementation. We will
 see this scheme results in production of photon pulses where each pulse has its own individualized
 single photon certification, and as expected, these certifications can be significantly better than the
 uniform result obtained from the conventional arrangement. To begin, we consider 
 the standard PDC setup for
 producing heralded single photons. The 
 trigger detector registers one photon of a pair and indicates the existence of the second photon exiting
 the correlated channel. The collection optics for that correlated channel 
 are designed to collect, as close
 as can be approximated, just the photons correlated to those seen by the trigger detector. In this 
arrangement, both the trigger channel and output channel are set up to collect only one mode of the 
field \cite{KOW01}. When the trigger detector fires, one photon has been received (assuming 
negligible dark counts), but we do not know if additional photons were also 
present as the considered detectors cannot distinguish a one-photon from a multi-photon event. 
Given that the trigger detector has fired, the probability that there were $n$ photons incident is:
\begin{equation}
P^{\text{F}}_{\overline{n},\eta}(n)= {(1-(1-\eta)^{n}) \times 
P_{\overline{n}}(n)\over \sum_{k=1}^\infty (1-(1-\eta)^{k}) \times 
P_{\overline{n}}(k)}%
\label{1}.
\end{equation}
Note that $1-(1-\eta)^{n}$ is the probability of the detector firing for 
$n$ photons incident and detector 
quantum efficiency, $\eta$ defined as the probability of the detector firing 
when just one photon is incident.
We use Bose statistics for the probability, $P_{\overline{n}}(n)$, 
of having $n$ photons emitted into a single mode of
 the PDC light, given $\overline{n}$ 
 \cite{ WAM95, GRT01}. 

With this basis, we now describe a system with a number of delay lines of increasing length placed
between the PDC crystal and the trigger detector. Each of these delay line channels intercepts
 a single, but separate, mode of the field. The output channel collection optic is also modified to include
 the extra modes correlated to those of the additional trigger modes.  Each
 of these trigger paths has a chance to cause the trigger detector to fire, with the shortest path 
providing the first chance, and the next longer path providing the next chance, and so on. But once the 
trigger detector fires due to a photon in a particular path, it cannot fire due to subsequent photons in
 the longer delay paths. The timing of the trigger detector firing relative to the pump pulse tells which
 delay path caused the firing. Thus the result of a single pulse of the pump laser is that either no 
trigger was produced or a trigger was produced and we know which delay path produced it. This last piece
 of information allows us to make a better determination of the probability that the photon produced
 was a single photon.  If the photon that causes the trigger to fire is one of the
 later delay paths we will have a much lower likelihood of there being more than one photon. We now 
calculate this likelihood as a function of which delay path caused the firing. 

For a system of $N_{\text{D}}$ delay paths where the $i^{\text{th}}$ delay path caused 
the firing, the one photon probability is: 
\begin{equation}
P_{\overline{n},\eta,N_{\text{D}}}(i)= 
 {\left({1-P^{\overline{\text{F}}}_{{\overline{n}\over{N_{\text{D}}}},\eta}}\right)^{i-1}}
P^{\text{F}}_{{\overline{n}\over{N_{\text{D}}}},\eta}(1)
\left({P_{\overline{n}\over{N_{\text{D}}}}(0)}\right)^{N_{\text{D}}-i}
\label{2},
\end{equation}
where $P^{\overline{\text{F}}}_{{\overline{n}\over{N_{\text{D}}}},\eta}$ is the probability that 
photons were incident if the detector did not fire.
Then we use Eq.~(\ref{1}) with $n$=1 
for a Bose distribution in each of the modes collected of Fig. 1b.
Figure 2 shows the functional form of these probabilities, where each line of the fan shaped family of
  curves represents a system of $N_{\text{D}}$ delays. Each point on a given line corresponds to 
the trigger firing at a particular $i^{\text{th}}$ delay out of a set of $N_{\text{D}}$. The point's value is the probability
 that this event 
indicates that exactly one photon pair exists in the system. It can be seen that for $N_{\text{D}}$ of more than 
a few, and with high $\eta$,
 we can have emission events with single photon probabilities around 90\%, greatly exceeding the conventional
 arrangement results for the same $\eta$ and $\overline{n}$. The conventional 
 result is 
represented by the single  $N_{\text{D}}$ =1 point (i.e. the standard PDC setup with one trigger path). 
	
We have calculated two additional probabilities: the overall probability of producing a 
heralded single photon (Eq.~(\ref{3})),
  and that same probability given that the trigger
 did fire (Eq.~(\ref{4})). The first is obtained by taking the product of the probability of a particular delay event
 occurring and the probability of that
 event being due to a single photon and then summing over all possible types of events. The
 second probability is obtained by eliminating the case where 
 the trigger did $not$ fire and renormalizing. 
Via some algebra we obtain:
\begin{equation}
{P_{1}}(\overline{n},\eta,N_{\text{D}})=
\overline{n} \eta \times
{\left({N_{\text{D}}\over{\overline{n}+N_{\text{D}}}}\right)^{1+N_{\text{D}}}}%
\label{3}
\end{equation}
\begin{equation}
{P_{1 
 }(\overline{n},\eta,N_{\text{D}}|trigger)}=  
{{{\overline{n} 
\eta\left({N_{\text{D}}\over{\overline{n}+N_{\text{D}}}}\right)^{1+N_{\text{D}}}}}
\over{1- {\left({N_{\text{D}}\over{\overline{n} \eta 
+N_{\text{D}}}}\right)^{N_{\text{D}}}}
}}%
\label{4}.
\end{equation}
for a Bose distribution in each of the modes collected in Fig. 1b.
These two results are also shown on Fig 2. For these curves, the abscissa 
is $N_{\text{D}}$  rather than delay $i$. Note that the total single photon probability rises somewhat 
  as $N_{\text{D}}\to \infty$. This probability,
 calculated for a thermal source approaches the result that would be obtained for a Poisson 
distribution $P_{1 }(\overline{n},\eta)=\overline{n} \eta 
~ e^{-\overline{n}}$, which is independent of $N_{\text{D}}$. This independence for a set of Poisson subsystems is what one 
would expect, as a collection of Poisson subsystems taken together yield a result with Poisson statistics. 
The tending of the collection of many single mode Bose subsystems toward the Poisson result is also understandable in
 that the more independent subsystems that are included in the sum, the more the individual events in the 
system are independent of each other, which is the definition of Poisson statistics. 
	The higher values of $P_{1 
 }(\overline{n},\eta,N_{\text{D}}|trigger)$  seen in Fig. 2 indicate the advantage of having heralded photons, 
rather than those from a faint laser. We also note that for Poisson  
instead of Bose distributions, the above analysis yields
 qualitatively the same shaped curves seen in Fig. 2, but all the 
 $P_{\overline{n},\eta,N_{\text{D}}}(i)$ probabilities are 
somewhat higher. 

 \begin{figure}
 \includegraphics{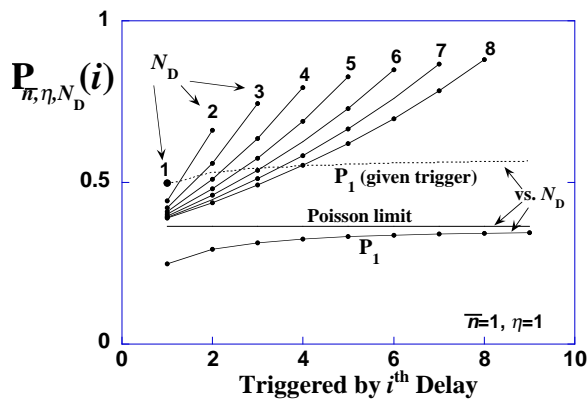}%
 \caption{\label{fig2} The fan of curves labeled 1-8 are the probabilities of exactly one photon 
being produced given that the $i^{\text{th}}$ delay, in a system of 
$N_{\text{D}}$ delays, caused the trigger to fire for $\eta$,$\overline{n}$=1.  The 
lowest curve is the total probability of a system of $N_{\text{D}}$ producing a single 
heralded photon per pump pulse with 
the Poisson limit shown just above. The dashed curve above the Poisson limit is the probability that 
the emitted light is a single photon given that the trigger did fire. (For these 
last 3 curves, the $x$-axis
 is $N_{\text{D}}$  rather than triggered delay.)}
 \end{figure}

Analysis of Eq.~(\ref{2}) shows that $\eta$  near one gives the best single photon certifications. 
This is because high $\eta$ means that the system provides more complete information about 
what has happened, e.g. an instance of the trigger not firing means with high 
certainty that no photon was incident, while low  $\eta$ decreases our certainty of that. Highlighting 
the advantage of this method, Fig. 2 also shows that in almost all cases, the multiplexed heralded system 
presented here significantly surpasses the single photon probability of a faint laser described 
by the Poisson limit to the curve. This advantage still holds for $\eta <$ 1, 
althought the certifications are not as high.

Eq.~(\ref{2}) also shows that increasing  $\overline{n}$ increases the spread of the certifications, while decreasing the maximum
 single photon certification possible. Thus there is a trade off between having high certainty
 single photons and high overall single photon number. In fact, $ P_{1 
 }$  does not continue increasing with increasing
 $\overline{n}$ . 
The maximum occurs for all systems at $\overline{n}$=1. This is the balancing 
point between reducing
 the number of single photon events and increasing the number of multiple photon 
events. This tells us the best rate to operate the system to maximize single photon events, although it will
 not necessarily provide the highest single photon certainties. We can still achieve higher single photon
 certifications than is possible conventionally, while maximizing single photon rates. 
	We must of course, verify that the events with high single photon probabilities 
	or ``certifications"
 have reasonable likelihoods of occurring. An analysis of this question shows 
 that while the likelihood of 
the later delay events occurring is lower than the earlier delay events, the dependence is not particularly
 strong. For instance, in the $N_{\text{D}}$=8, $\overline{n}$=1 case the falloff from delay 1 events 
to delay 8 events is only a factor of
 about 2. 

We have shown a way to decouple the probabilities of producing a single photon and the probability
 of producing more than one photon using an array of parametric downconverters.  By doing so, we can 
construct a better approximation of a true single photon on demand source than is possible using a 
conventional single PDC setup and certainly better than a faint laser 
source \cite{BLM00, GRT01}. In principle this method could achieve an arbitrarily close
 approximation of a single photon on demand source. We have also analyzed a version 
 that, while
greatly simplifying the construction of an actual device, retains most of the benefits of the 
original concept. We have shown that the setup would produce single photons with individual certifications
 that the photons produced are actually the desired single photons. Such a 
 better-defined single photon 
source will allow for better use of quantum channel resources in a cryptographic system by reducing the need 
for overhead tasks such as privacy amplification, as well as having implications for the field of quantum 
computation. As photon counting becomes more convenient at telecom wavelengths we expect that integrated 
all solid state implementations of these schemes will be made even easier and we will have truly achieved
 the dream of a convenient source of single photons on demand. We are currently working on experimental 
implementations. 

\end{document}